\documentclass[aps,prl,reprint,superscriptaddress]{revtex4-2}
\usepackage{amsmath}
\usepackage{graphicx}
\input{epsf}
\begin{document}
	%Title of paper
	\title{A thermal model with AC Josephson effect for a shunted superconducting weak-link}
	\author{Sourav Biswas}
	\email[]{sbiswas.physics@gmail.com}
	%\homepage[]{Your web page}
	%\thanks{}
	\affiliation{Department of Physics, Indian Institute of Technology Kanpur, Kanpur 208016, India}
%	\affiliation{Department of Physics, Indian Institute of Technology Kanpur, Kanpur 208016, India}
%	%\author{Nikhil Kumar}
%	%\affiliation{Department of Physics, Indian Institute of Technology Kanpur, Kanpur 208016, India}
%	\author{Clemens B. Winkelmann}
	\affiliation{Braun Center for Submicron Research, Department of Condensed Matter Physics, Weizmann Institute of Science, Rehovot 7610001, Israel}
%	%\affiliation{CNRS, Institut N\'eel, F-38042 Grenoble, France}
%	\author{Herv\'e Courtois}
%	\affiliation{\mbox{Univ.} Grenoble Alpes, CNRS, Grenoble INP, Institut N\'eel, Grenoble, France}
%	%\affiliation{CNRS, Institut N\'eel, F-38042 Grenoble, France}
	\author{Anjan Kumar Gupta}
	\affiliation{Department of Physics, Indian Institute of Technology Kanpur, Kanpur 208016, India}
%	\date{\today}
	
	\begin{abstract}
Superconducting weak-link (WL), behaving like a Josephson junction (JJ), is fundamental to many superconducting devices such as nanoSQUIDs, single-photon detectors, and bolometers. The interplay between unique nonlinear dynamics and inevitable Joule heating in a JJ leads to new characteristics. Here, we report a time-dependent model incorporating thermal effect in the AC Josephson regime for a Josephson WL shunted by a resistor together with an inductor to investigate the dynamics as well as the resulting current-voltage characteristics. We find that the dynamic regime where phase and temperature oscillate simply widens due to a pure resistive shunt. However, a significant inductive time-scale in the shunt loop, competing with the thermal time-scale, introduces high-frequency relaxation oscillations in the dynamic regime. Based on numerical analysis, we present state diagrams for different parameter regimes. Our model is a guide for better controlling the parameters in the experiments of WL-based devices.
	\end{abstract}
	
	\maketitle
\section{Introduction}	
Josephson junction (JJ) is one of the excellent candidates in device physics for its diverse application from superconducting quantum interference device (SQUID) \cite{squidbook,tinkham book} to quantum computation \cite{quantum}. In recent years, superconducting weak-link (WL) or nanowire-based JJ has been, in particular, a major focus for nano(n)SQUID magnetometry, single-photon detection, nanowire bolometer, etc. \cite{likharev,granatasquid,recentsquidreview}. The most interesting state of a JJ is the phase dynamic state where the superconducting phase oscillates according to the AC Josephson effect. However, unlike conventional JJ, a WLs' phase actively plays with its raised temperature due to Joule heating in the junction. This gives rise to a different transport behavior \cite{tinkham,herve,anjanjap} than the one expected from the widely used resistively and capacitively shunted junction (RCSJ) model \cite{squidbook}. A common phenomenon is the hysteretic current-voltage characteristics (IVCs) with a critical current ($I_{\rm c}$) and a retrapping current $I_{\rm r}$ ($< I_{\rm c}$), but no phase coherence in the dissipative state due to overheating \cite{skocpol,hazra,nikhil prl,nayana}. A hysteretic JJ exhibiting sharp switching at $I_{\rm c}$ has helped to develop efficient single-photon detectors \cite{singlephoton,singlephoton-2,singlephoton-new}.

A hysteretic WL with moderate heating and/or small $I_{\rm c}$ can have a non-zero phase coherence (Josephson coupling) coexisting with dissipation with $T_{\rm c}>T_{\rm WL}>T_{\rm b}$; here $T_{\rm WL}$ ($T_{\rm b}$) is the WL (bath) temperature \cite{anjanjap,sourav}. This mixed regime is obligatory for a WL-based nSQUIDs' employment as a flux-to-voltage transducer \cite{nanomag1,nanomag2,nanomag3}. However, a much better flux sensitivity is obtained when the WLs are non-hysteretic at a higher $T_{\rm b}$, close to $T_{\rm c}$. Several methods for reducing thermal hysteresis and increasing nSQUIDs' sensitivity were demonstrated \cite{lam,dibyendu,nikhil sust,souravRL,sagarpra}. A very effective technique was a shunt resistor and an inductor parallel to the WL \cite{nikhil sust,souravRL}; although a large inductance in the shunt loop was seen to produce relaxation oscillations, reported on Nb and Sn micro-bridges \cite{skocpol,muck}.

Relaxation oscillations were first observed in resistively shunted point contact SQUID by Zimmerman et al. \cite{zimmerman}. The observations were explained by Taur et al. with a simple circuit model for a point contact JJ \cite{taur}. Later, several groups studied the properties of such oscillations in resistively and inductively shunted Josephson tunnel junction \cite{vernon,relaxapl,relaxiop}. Inductive effect on the RCSJ model and the observed characteristics in relaxation mode were discussed elsewhere \cite{sullivan,whan}. These studies showed another perspective of a JJ as a relaxation oscillator, which can produce very high-frequency oscillations up to the order of GHz \cite{calander,kerl berggren mit,hadfield}. High power generation of such oscillations makes the JJ relaxation oscillator useful in some applications \cite{relax-application1,relax-application2}.

For a WL-type superconducting structure, an RSJ model (not an RCSJ model) with heat balance provides an appropriate understanding. Therefore, a detailed quantitative understanding of the dynamics of a shunted WL with respect to the relevant parameters in the system will be useful for operating such devices in different regimes. In this work, we consider a superconducting WL shunted by a parallel resistor and an inductor, and study the temperature and phase dynamics using a time-dependent thermal model. Finding various states for different parameter regimes, we describe the state diagrams and IVCs of the device. In a limiting case, the characteristics of a bare WL are obtained. The effect of non-sinusoidal supercurrent-phase relation in longer WL limit is also discussed.

\section{The Model}
\begin{figure}[h!]
	\centering
	\includegraphics[width=2.0in]{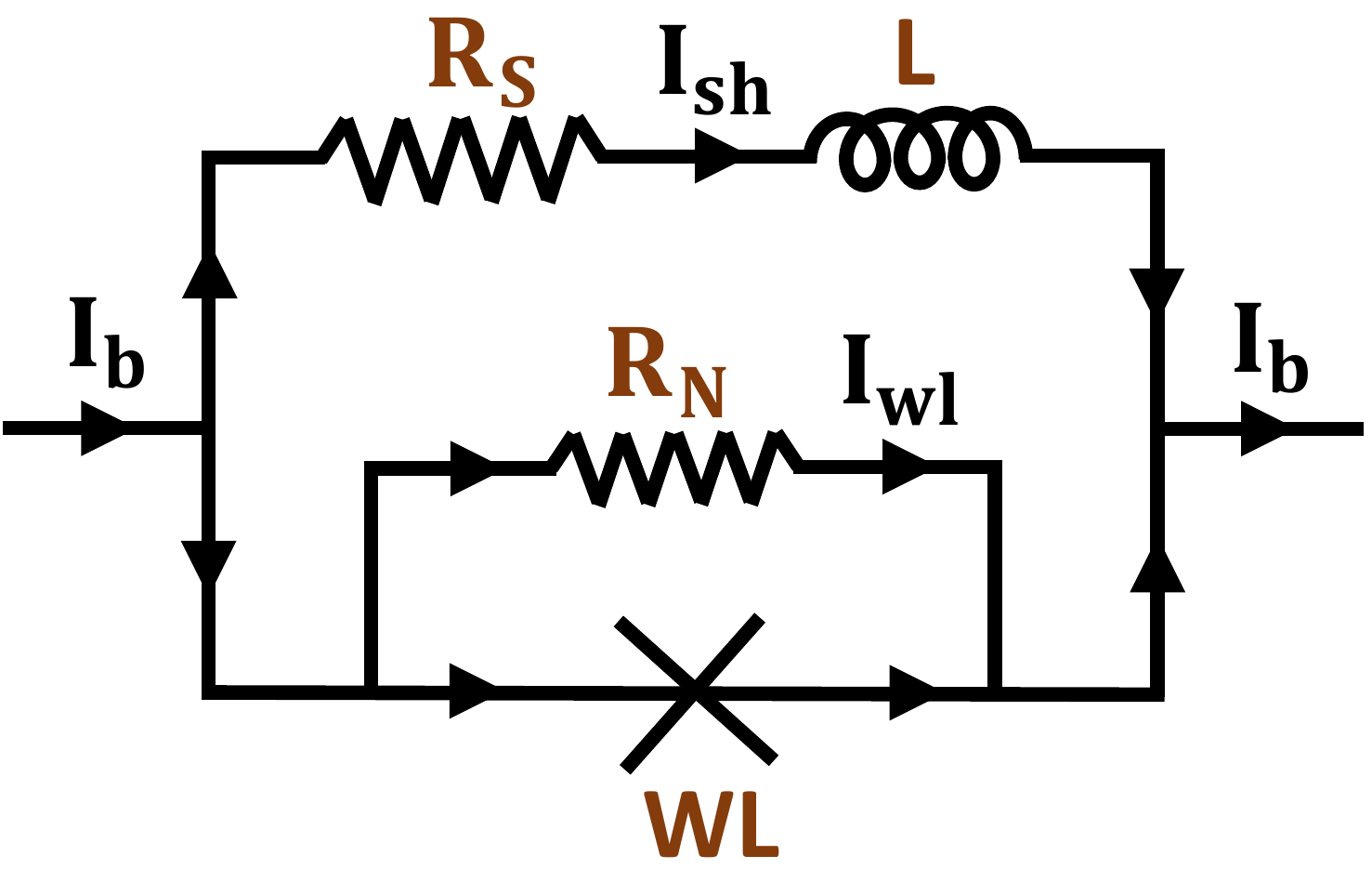}
	\caption{Schematic diagram of a current biased resistively and inductively shunted superconducting weak-link (WL).}
	\label{fig1}
\end{figure}

Figure \ref{fig1} shows the circuit diagram of a current biased superconducting WL with a parallel shunt resistor $R_{\rm S}$ and the loop inductance $L$. The total bias current $I_{\rm b}$ is shared between the shunt path as $I_{\rm sh}$, and the WL path. The latter part is also dynamically shared between the supercurrent $I_{\rm sc}$ and normal current $I_{\rm wl}$ (due to WLs' normal resistance $R_{\rm N}$) of the WL. Supercurrent $I_{\rm sc}$ is defined by the phase difference $\varphi$ of the two bulk superconductors across the WL \cite{tinkham book}. Depending on the WL dimension, $I_{\rm sc}$ evolves differently with $\varphi$. Considering a general $\varphi$ dependence $g(\varphi)$ of the supercurrent, $I_{\rm sc}=I_{\rm c}(T)g(\varphi)$ with $I_{\rm c}$ as the maximum supercurrent or the critical current, we have
\begin{equation}
	I_{\rm b}=I_{\rm sh}+I_{\rm c}(T)g(\varphi)+\frac{V}{R_{\rm N}}.
\end{equation}
The phase difference $\varphi$ is given by the AC Josephson relation $V=(\hbar/2e)d\varphi/dt=(\Phi_{\rm 0}/2\pi)d\varphi/dt$ with $\Phi_{\rm 0}=h/2e$, the flux quantum ($h$ is the Planck constant and $e$ the electron charge).

The Joule power ($P=VI$) in the WL is given by $P=V[I_{\rm c}(T)g(\varphi)+I_{\rm wl}]=[\frac{\hbar I_{\rm c}(T)}{2e}\dot{\varphi}g(\varphi)+\frac{\hbar^2}{4e^2R_{\rm N}}\dot{\varphi}^2]$. The first term $\textendash$ rate of change in the Josephson energy $\textendash$ being proportional to $\sin(\varphi)$, gives zero contribution to the average $P$ over a full $2\pi$ cycle in $\varphi$ (see also texts later). The second term, \mbox{i.e.}, the Joule power dissipation $V^2/R_{\rm N}$ due to a normal current $V/R_{\rm N}$, only provides a nonzero average value leading to the WL getting hot. Thus, omitting the Josephson energy related-term, the heat balance in the WL is governed by
\begin{equation}
	C_{\rm WL}\frac{dT}{dt}+ k ( T - T_{\rm b} ) =\frac{V^{2}}{R_{\rm N}}=\frac{(\Phi_{\rm 0})^{2}}{4\pi^{2}R_{\rm N}}\Big(\frac{d\varphi}{dt}\Big)^{2}.
\end{equation}
Here, $C_{\rm WL}$ and $k$ are temperature-independent heat capacity and heat loss coefficient, respectively. $T$ and $T_{\rm b}$ are the instantaneous WL temperature and the bath temperature, respectively. We assume the WL dimension to be less than the thermal healing length \cite{skocpol,nikhil prl} so that the whole WL is considered at a single temperature $T$. Current ($I_{\rm sh}$) through the shunt resistor $R_{\rm s}$ and inductor $L$ gives rise to an additional equation given by
\begin{equation}
	L\frac{dI_{\rm sh}}{dt}+I_{\rm sh}R_{\rm S}=V=\frac{\Phi_{\rm 0}}{2\pi}\frac{d\varphi}{dt}.
\end{equation}
Simplification of these three equations in terms of dimensionless variables results into the following forms:
\begin{equation}
	\dot{\varphi}=2\pi[i_{\rm b}-i_{\rm c}(p)g(\varphi)-i_{\rm sh}]
	\label{eq:eq1}
\end{equation}
\begin{equation}
	\alpha \dot{p}+p=\frac{\beta}{4\pi^2}\dot{\varphi}^{2}
	\label{eq:eq2}
\end{equation}
\begin{equation}
	\gamma\dot{i_{\rm sh}}+i_{\rm sh}=\frac{r}{2\pi}\dot{\varphi}.
	\label{eq:eq3}
\end{equation}
Here, $\gamma=\tau_{\rm L}/\tau_{\rm J}$, $\alpha = \tau_{\rm th}/\tau_{\rm J}$, inductive time-scale $\tau_{\rm L}=L/R_{\rm S}$, thermal time-scale $\tau_{\rm th} = C_{\rm WL}/k$, $r=R_{\rm N}/R_{\rm S}$, and $p = (T- T_{\rm b})/(T_{\rm c}-T_{\rm b})$ is the dimensionless temperature. Also, all currents are normalized using $I_{\rm c}(T_{\rm b})$, e.g. $i_{\rm b} = I_{\rm b}/I_{\rm c}(T_{\rm b})$ and the dimensionless time variable is $\tau = t/\tau_{\rm J}$ with $\tau_{\rm J}$ the Josephson time scale, defined as $\tau_{\rm J}=\frac{\Phi_{\rm 0}}{R_{\rm N}I_{\rm c}(T_{\rm b})}$. Bath temperature dependent $\beta =\frac{I_{\rm c}^2(T_{\rm b})R_{\rm N}}{k(T_{\rm c}-T_{\rm b})}$ determines the heat generation to evacuation ratio in the junction.

\section{Results and Discussion}
\subsection{\textbf{Static steady states}}
There are two types of steady states possible for Eqs. (\ref{eq:eq1}-\ref{eq:eq3}): (i) Static steady state where temperature and phase do not have any time dependence and (ii) dynamic steady state in which both of them oscillate. These two steady states together determine the DC IVCs of the device.

The static steady state corresponding to $\dot{p}=0$ and $\dot{\varphi}=0$ (for $p<1$) occurs when shunt current $i_{\rm sh}$ is zero and full bias current $i_{\rm b}$ flows through the WL, \mbox{i.e.}, $i_{\rm b}=i_{\rm c}(p)g(\varphi)$ and $T=T_{\rm b}$. The other static steady state exists with $p>1$ or $T>T_{\rm c}$; thus the WL is normal with no supercurrent left, \mbox{i.e.}, $i_{\rm c}(p)=0$. The bias current $i_{\rm b}$ is then shared between the two resistors $R_{\rm N}$ and $R_{\rm S}$ as WLs' normal current $i_{\rm wl}$ and shunt current $i_{\rm sh}$, respectively, \mbox{i.e.}, $i_{\rm b}=i_{\rm wl}+i_{\rm sh}$. Eliminating $\dot{\varphi}$ from  Eqs. (\ref{eq:eq1}) and (\ref{eq:eq2}), we get $p=\beta (i_{\rm b}-i_{\rm sh})^2=\beta i^2_{\rm wl}=\frac{\beta i^2_{\rm b}}{(1+r)^{2}}$. This provides the limit (when $p=1$) of bias current, called the static retrapping current, to be
\begin{equation}
	i_{\rm sr}=\frac{1+r}{\sqrt{\beta}} \hspace{0.5cm} [Or, \hspace{0.2cm}I_{\rm sr}=\sqrt{\frac{\kappa (T_{\rm c}-T_{\rm b})}{R_{\rm N}}}(1+r)].
\end{equation}
Above $I_{\rm sr}$, the WL becomes fully normal (resistive). Note that the two limits of the static steady states are not affected by the loop inductance $L$; only the value of shunt resistance $R_{\rm S}$ alters the value of $I_{\rm sr}$.

\subsection{\textbf{Dynamic steady states}}
\textit{i. For negligible $\gamma$:} The characteristics of the dynamic (oscillatory) steady-state over the current range between the fully superconducting and fully normal states depends on the value of $\tau_{\rm L}/\tau_{\rm th}=\gamma/\alpha$. For very small values of $\tau_{L}$, neglecting $\gamma$ and eliminating $i_{\rm sh}$ from  Eqs. (\ref{eq:eq1}) and (\ref{eq:eq3}), we obtain
\begin{equation}
	\dot{\varphi}=\frac{2\pi}{1+r}\{i_{\rm b}-i_{\rm c}(p)g(\varphi)\}.
	\label{eq:eq4}
\end{equation}
Integrating Eq. (\ref{eq:eq4}) when $\varphi$ changes by $2\pi$ for one phase-slip process \cite{tinkham book}, the corresponding time scale $\tau_{\rm ps}$ is obtained as
\begin{equation}
	\tau_{\rm ps}=\frac{1+r}{2\pi}\int_{0}^{2\pi}\frac{d\varphi}{\big[i_{\rm b}-i_{\rm c}(p)g(\varphi)\big]}.
	\label{eq:taups}
\end{equation}
Also, $\langle\dot{\varphi}^2\rangle_{\tau_{\rm ps}}$ can be written as $\langle\dot{\varphi}^2\rangle_{\tau_{\rm ps}}=\frac{1}{\tau_{\rm ps}}\int_{0}^{2\pi}\dot{\varphi}d\varphi$, or
\begin{equation}
	\langle\dot{\varphi}^2\rangle_{\tau_{\rm ps}}
	=\frac{2\pi}{(1+r)\tau_{\rm ps}}\int_{0}^{2\pi}\big[i_{\rm b}-i_{\rm c}(p)g(\varphi)\big]d\varphi.
	\label{eq:six}
\end{equation}
For a short WL, \mbox{i.e.}, when the WLs' length $\ell$ is much shorter than the superconducting coherence length $\xi$ ($\ell/\xi\ll1$), $g(\varphi)$ takes the form: $g(\varphi)=\sin(\varphi)$. With this, the Eqs. (\ref{eq:taups}) and (\ref{eq:six}) get simplified into $\tau_{\rm ps}=\frac{1+r}{\sqrt{i_{\rm b}^2-i_{\rm c}^2(p)}}$ and $\langle\dot{\varphi}^2\rangle_{\tau_{\rm ps}}=\frac{4\pi^2i_{\rm b}}{(1+r)\tau_{\rm ps}}$, respectively.

To obtain the temperature dynamics over one phase-slip time, we write Eq. \ref{eq:eq2} as $\alpha \langle\dot{p}\rangle_{\tau_{\rm ps}}=-\langle p\rangle+\frac{\beta i_{\rm b}}{(1+r)\tau_{\rm ps}}$. This can be written in terms of a fictitious potential ($U$) as $dU/dp=-p+\frac{\beta i_{\rm b}}{(1+r)^2}{\sqrt{i_{\rm b}^2-i_{\rm c}^2(p)}}$. For $i_{\rm b}\leq i_{\rm c}(p)$, $dU/dp=-p$ corresponds to the $U(p)$ minimum at $p=0$ ($\mbox{i.e.}, T=T_{\rm b}$), the superconducting state. The other minimum of $U(p)$ with $0<p<1$ for $i_{\rm b}>i_{\rm c}(p)$ can be found by solving $dU/dp=0$. This provides a (minimum) condition on $i_{\rm b}$ for holding the $U(p)$ minimum at $0<p<1$ ($\mbox{i.e.}, T_{\rm b}<T<T_{\rm c}$). Therefore, a minimum bias current which is called the dynamic retrapping current $i_{\rm dr}$ \cite{anjanjap}, is required to sustain the steady dynamic regime with WL temperature above $T_{\rm b}$ but below $T_{\rm c}$. Assuming linear temperature dependence of $i_{\rm c}(p)$, \mbox{i.e.}, $i_{\rm c}(p)=(1-p)\Theta(1-p)$, we find the expression for $i_{\rm dr}$ to be,
\begin{equation} i_{\rm dr}^{2}=\frac{\sqrt{1+\frac{4\beta^{2}}{(1+r)^{4}}}-1}{2\beta^{2}/(1+r)^{4}}.
	\label{eq:eq5}
\end{equation}
\begin{figure}[h!]
	\includegraphics[width=0.9\columnwidth]{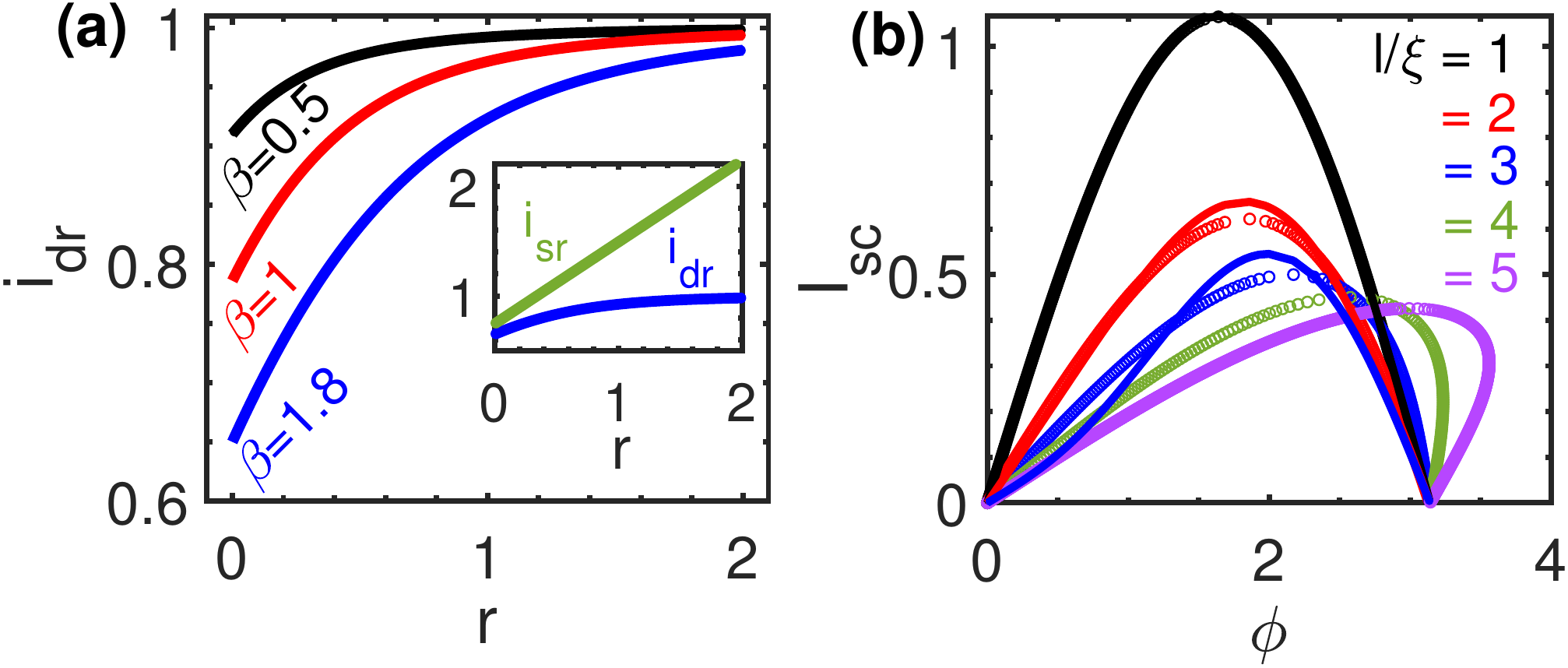}
	\centering
	\caption{(a) Variation of dynamic retrapping current ($i_{\rm dr}$) with $r=\frac{R_{\rm N}}{R_{\rm S}}$, neglecting the effect of $\gamma$ for small $\tau_{\rm L}$. Inset displays $i_{\rm dr}$ along with static retrapping current ($i_{\rm sr}$) as a function of $r$ for $\beta=1.8$. (b) Non-sinusoidal behavior of supercurrent-phase ($I_{\rm sc}-\varphi$) relation for different $\ell/\xi$ values. Solid lines are the fits to the Eq. (\ref{eq:fit}). The fitting parameters are $I_{\rm c}=1.067, C=0.034$ (for $\ell/\xi=1$); $I_{\rm c}=0.643, C=0.1425$ (for $\ell/\xi=2$) and $I_{\rm c}=0.5, C=0.306$ (for $\ell/\xi=3$). $I_{\rm sc}$ is multivalued for $\ell/\xi\geq4$.}
	\label{fig2}
\end{figure}

	Note that for a bare resistive shunt with a WL, effective $\beta$ gets modified to $\beta/(1+r)^2$, which has a drastic effect on the dynamic regime. Variation of $i_{\rm dr}$ with $r$ for different values of $\beta$ is shown in Fig. \ref{fig2}(a). As $r$ increases, \mbox{i.e.}, shunt resistance $R_{\rm S}$ decreases with respect to the WLs' normal resistance $R_{\rm N}$, more normal current passes through $R_{\rm S}$ when voltage generation occurs due to phase-slip processes. Thus, the WL requires a larger bias current to maintain steady oscillations in temperature $p$ below $T_{\rm c}$; thus an increased $i_{\rm dr}$ with $r$. In another way, $\beta$ is interpreted as the ratio of heat generation to heat evacuation; thus a higher value of $\beta$ leads to a lower $i_{\rm dr}$ for a particular $r$. In the Fig. \ref{fig2}(a) inset, we show the widening of the dynamic regime between the dynamic retrapping current $i_{\rm dr}$ and the static retrapping current $i_{\rm sr}$ as a function of $r$ at a $\beta=1.8$. The dynamic regime has a finite supercurrent and thus useful for probing voltage modulations with magnetic field in a nSQUID. This was recently demonstrated for an unshunted nSQUID, \mbox{i.e.}, when $r=0$ \cite{sourav}.

What would happen if WLs' length is comparable to or longer than the coherence length $\xi$? In this case, the sinusoidal supercurrent-phase ($I_{\rm sc}-\varphi$) relation is no longer valid. Numerical solution of Ginzburg-Landau (GL) equation \cite{likharev,wolf,golubov} for a long WL gives rise to a non-sinusoidal variation of $I_{\rm sc}$ with $\varphi$, see Fig. \ref{fig2}(b). These $I_{\rm sc}-\varphi$ curves can be described using a general equation,
\begin{equation}
	I_{\rm sc}=\frac{I_{\rm c}}{\sqrt{1+C^2}}\big[\sin(\varphi)-C\sin(2\varphi)\big],
	\label{eq:fit}
\end{equation}
where $C$ is a constant factor; $C=0$ makes it ideal Josephson junction. The fit to Eq. \ref{eq:fit} for $\ell/\xi=1$ to $3$ is shown in Fig. \ref{fig2}(b). Putting $g(\varphi)$ thus obtained, in Eq. \ref{eq:taups} and Eq. \ref{eq:six}, we find no simplified form of $\tau_{\rm ps}$ like the one for sinusoidal case. Therefore, we estimate the numerical values of $\tau_{\rm ps}$ for $I_{\rm sc}-\varphi$ relation till $\ell/\xi=3$ and find very minute change in the value of $\tau_{\rm ps}$ (the values are found to be the same up to three decimal places) as compared to that for sinusoidal one. This indicates that the WLs' length and hence the nature of $I_{\rm sc}-\varphi$ within our accessible regime has negligible effect on the dynamic retrapping current, $i_{\rm dr}$. This is further confirmed from the almost alike IVCs for different $g(\varphi)$, shown later.

\begin{figure}[h!]
	\centering
	\includegraphics[width=0.85\columnwidth]{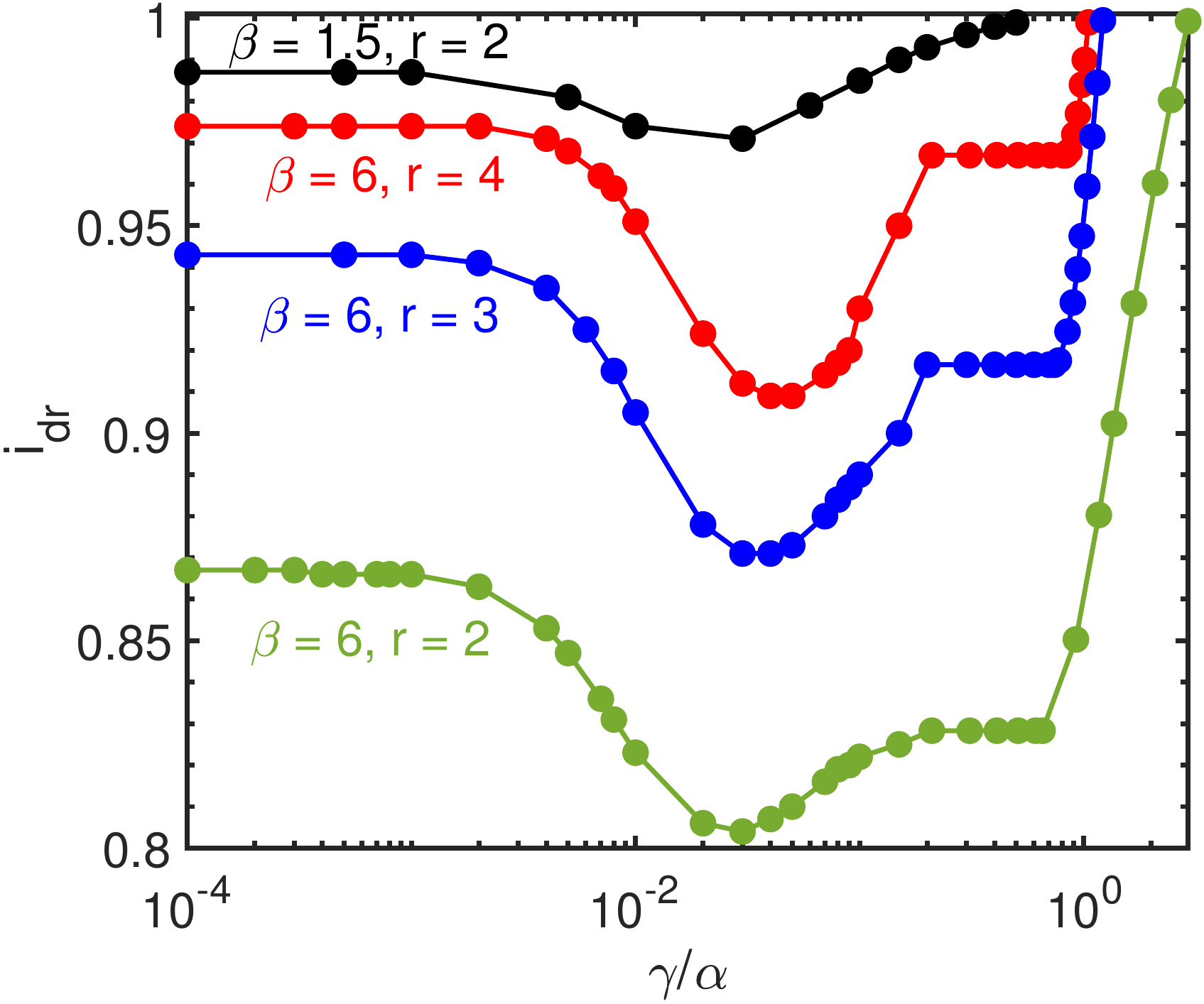}
	\caption{Effect of $\gamma/\alpha$ on dynamic retrapping current $i_{\rm dr}$ for few $\beta$ and $r$ values. Solution is done with $\alpha$=100.}
	\label{fig3}
\end{figure}
\textit{ii. For finite $\gamma$:} We now look for the effect of $\gamma$ on the dynamic regime for short WL limit only with $g(\varphi)=\sin(\varphi)$. We solve the Eqs. (\ref{eq:eq1}-\ref{eq:eq3}) numerically by tuning $\gamma$ and finding the minimum $i_{\rm b}$ as $i_{\rm dr}$, at which phase starts appearing dynamic. Figure \ref{fig3} shows the variation of thus-obtained $i_{\rm dr}$ with $\gamma/\alpha$ for few sets of $\beta$ and $r$. When $\gamma\ll1$ (or, more precisely $\gamma\ll\alpha$), as expected $i_{\rm dr}$ is $\gamma$ independent and given by the Eq. (\ref{eq:eq5}). With increased $\gamma$ while other parameters are fixed, $i_{\rm dr}$ is found to shift from the plateau and ultimately move towards $1$ following a pattern, see Fig. \ref{fig3}. In the intermediate range, the hysteresis gets wider with a $i_{\rm dr}$-minimum (see also IVCs later), which can be benefical for Josephson switching study. This intermediate $i_{\rm dr}$ variation with $\gamma/\alpha$ is not intuitively clear to us. A report \cite{whan_chaos} showed some chaotic behavior also over the intermediate range of $\gamma$ ($\mbox {i.e.}$, when $\tau_{\rm L}$ and $\tau_{\rm th}$ are comparable). However, the subsequent plateau and the variation near 1 can be understood using nonlinear analysis, see our Ref. \cite{souravRL}. This particular regime has remarkable impact on a practical nSQUID operation as it leads to non-hysteretic characteristic with striking SQUID-sensitivity \cite{souravRL,sagarnano}.
%	\onecolumngrid
%	\twocolumngrid	

\begin{figure}
	\centering
	\includegraphics[width=\columnwidth]{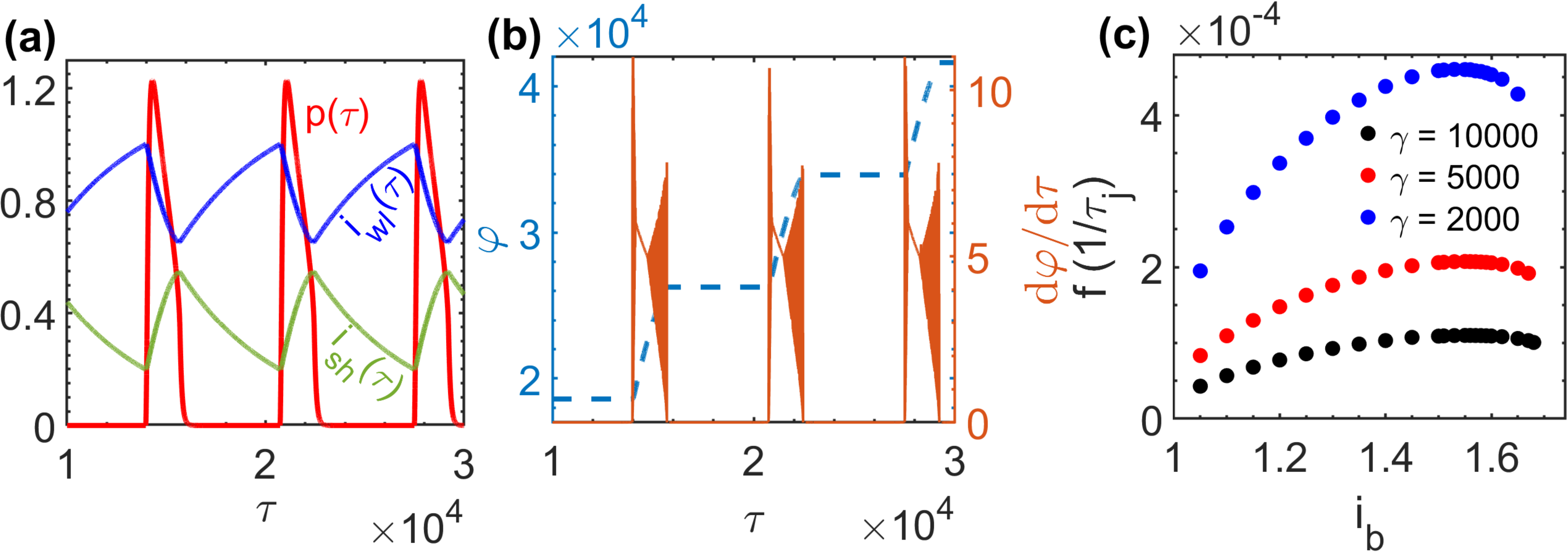}
	\caption{(a) Relaxation oscillations in temperature ($p$), shunt current ($i_{\rm sh}$) and WLs' normal current ($i_{\rm wl}$) at bias $i_{\rm b}=1.2$ for the parameters $\beta=1.5, r=2$ and $\gamma=5000$. (b) The oscillation in phase derivative ($\dot{\varphi}$) and the variation of phase ($\varphi$). (c) Frequency ($f$) of relaxation oscillations with bias current ($i_{\rm b}$) at three $\gamma$ values.}
	\label{fig4}
\end{figure}

As $\gamma$ (or $\tau_{\rm L}$) is increased further beyond its limit at which $i_{\rm dr}$ reaches $1$, the behavior of the dynamic state changes, which we discuss next for $r=2$ and $\beta=1.5$. First, we simplify the Eqs. (\ref{eq:eq1}-\ref{eq:eq3}) in standard fashion like before. $\tau_{\rm ps}$ is now given by $\tau_{\rm ps}=\frac{1}{\sqrt{(i_{\rm b}-i_{\rm sh})^2-i^2_{\rm c}(p)}}$. $\langle\dot{\varphi}^2\rangle_{\tau_{\rm ps}}$ and $\langle\dot{\varphi}\rangle_{\tau_{\rm ps}}$ come out as $\frac{4\pi^2(i_{\rm b}-i_{\rm sh})}{\tau_{\rm ps}}$ and $\frac{2\pi}{\tau_{\rm ps}}$, respectively. From the Eqs. (\ref{eq:eq2}) and (\ref{eq:eq3}), we thus obtain over a single phase-slip
\begin{equation}
	\gamma\dot{i_{\rm sh}}+i_{\rm sh}=r\sqrt{(i_{\rm b}-i_{\rm sh})^2-i^2_{\rm c}(p)},
	\label{eq:eq6}
\end{equation}
\begin{equation}
	\alpha\dot{p}+p=\beta(i_{\rm b}-i_{\rm sh})\sqrt{(i_{\rm b}-i_{\rm sh})^2-i^2_{\rm c}(p)}.
	\label{eq:eq7}
\end{equation}
However, these equations are not analytically solvable except in certain time spans when $\dot{\varphi}=0$ or $i-i_{\rm sh}=i_{\rm c}(p)$. They then simply describe the exponential decay of $p$ and $i_{\rm sh}$. Therefore, to obtain the continuous time variation in temperature ($p$), shunt current ($i_{\rm sh}$) and WLs' normal current ($i_{\rm wl}$), we numerically solve these equations with $i_{\rm c}(p)=(1-p)\Theta(1-p)$ and $\alpha=100$ which is a good choice \cite{anjanjap,skocpol}. Steady state solutions of $p$, $i_{\rm wl}$ and $i_{\rm sh}$ for $\gamma=5000, \beta=1.5$ and $r=2$ are displayed in Fig. \ref{fig4}(a). Figure \ref{fig4}(b) represents the same for $\dot{\varphi}$ and $\varphi$. Robust relaxation oscillations in temperature, phase and currents are observed. Note that the WLs' state oscillates between fully superconducting ($p=0$) and fully normal ($p\geq1$).

We qualitatively describe this behavior as follows. When $i_{\rm b}=1$, i.e., $I=I_{\rm c}(T_{\rm b})$ at a time $\tau=\tau_{\rm 0}$, $\dot{\varphi}$ rises from zero intrinsically leading to AC Josephson effect. The JJ becomes resistive, and some current gets directed towards shunt. However, due to large $\tau_{\rm L}$ (or $\gamma$), shunt current ($i_{\rm sh}$) rises exponentially while the normal current ($i_{\rm wl}$) through $R_{\rm N}$ decays in the same manner. Since $i_{\rm wl}$ decreases, the temperature $p$ also decays after reaching a sudden peak due to initial jump in WL-state. The exponential decay (rise) of $i_{\rm wl}$ ($i_{\rm sh}$) and the dynamics of $p$ for non-zero $\dot{\varphi}$ are dictated by the Eqs. (\ref{eq:eq6}) and (\ref{eq:eq7}) with a boundary condition that at $\tau=\tau_{\rm 1}$, $i_{\rm wl}(\tau_{\rm 1})=i-i_{\rm sh}(\tau_{\rm 1})=1-p(\tau_{\rm 1})$. The other boundary condition is $i_{\rm sh}(\tau_{\rm 0})=p(\tau_{\rm 0})=0$. Therefore, at $\tau_{\rm 1}$, $i_{\rm wl}$ crosses $i_{\rm c}(p)$, which makes the device superconducting again with $\dot{\varphi}$ abruptly dropping to zero and $\varphi$ being constant. Consequently, $p$ diminishes to zero exponentially with thermal time constant $\alpha$, while $i_{\rm sh}$ ($i_{\rm wl}$) decreases (increases) exponentially with time constant $\gamma$. This continues till $\tau=\tau_{\rm 2}$ when $i_{\rm wl}$ reaches $1$ again making WL resistive with non-zero $\dot{\varphi}$. Thus, the steady relaxation cycles in the system continue with frequency $1/(\tau_{\rm 2}-\tau_{\rm 0})$. The oscillation frequency with bias current $i_{\rm b}$ is plotted in Fig. \ref{fig4}(c) for three different $\gamma$ values. For a typical value of $\tau_{\rm j}\approx10^{-12}$ sec, frequency is estimated to be of the order of MHz  \cite{kerl berggren mit}.

\begin{figure}[h!]
	\centering
	\includegraphics[width=\columnwidth]{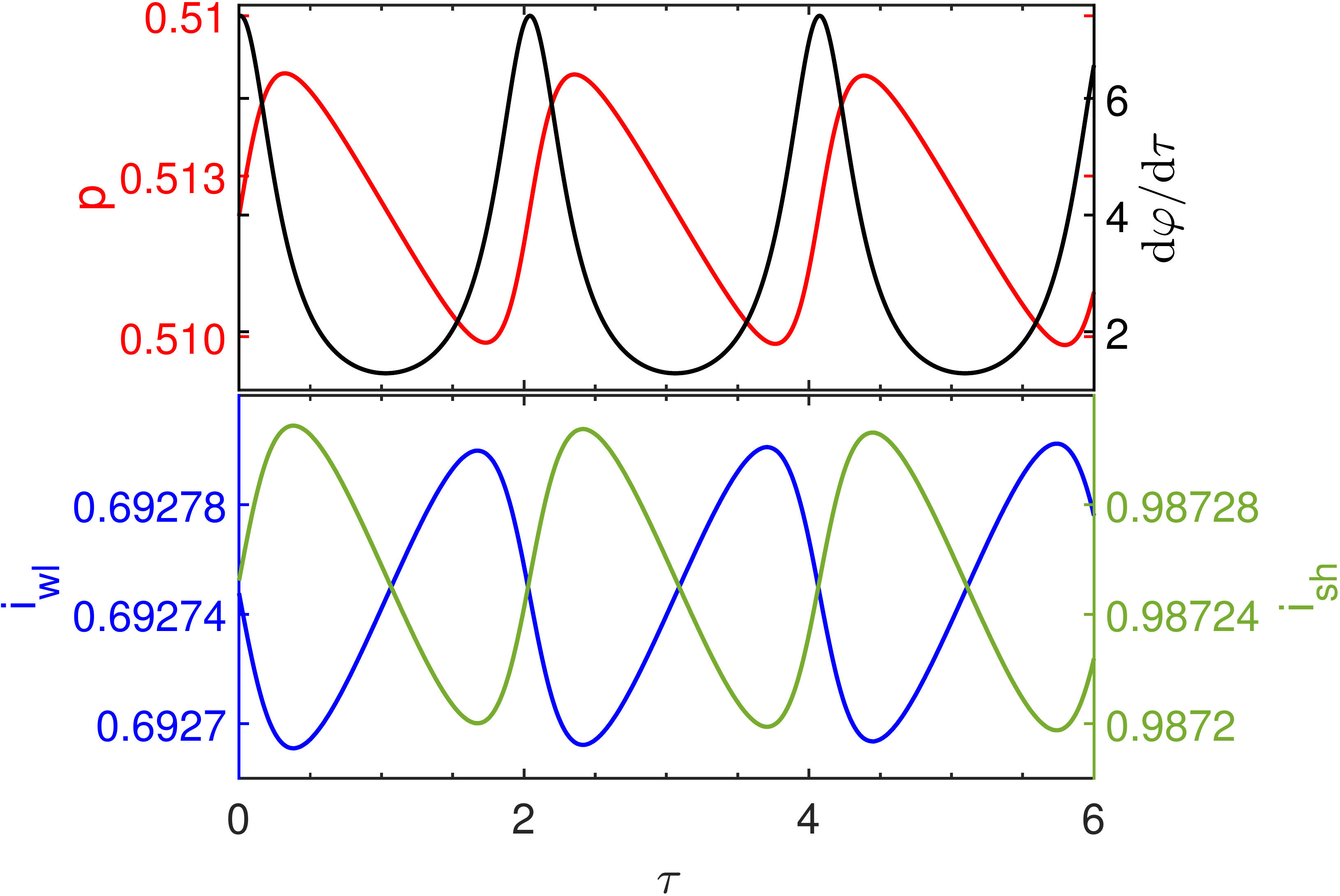}
	\caption{Dynamic steady state with small oscillation in (top) temperature $p$ and phase derivative $\dot{\varphi}$, and (bottom) WLs' normal current $i_{\rm wl}$ and shunt current $i_{\rm sh}$ at a bias current $i_{\rm b}=1.68$ for the parameters $\beta=1.5, r=2$ and $\gamma=5000$.}
	\label{fig5}
\end{figure}	
As the relaxation frequency increases with $i_{\rm b}$, we find a certain limit (say $i_{\rm max}$) of $i_{\rm b}$ beyond which $i_{\rm wl}$ becomes unable to cross $i_{\rm c}(p)$, instead it reaches a saturated steady small oscillations similar to the one with negligible inductance. For $\gamma=5000, \beta=1.5$ and $r=2$, $i_{\rm max}$ is found to be 1.679. Near the $i_{\rm b}=i_{\rm max}$, $f$ has a tendency to slightly reduce, see higher $i_{\rm b}$ region in Fig. \ref{fig4}c. This is possibly due to the crossover from relaxation state of the device. Above $i_{\rm max}$, the temperature stabilizes to a value below $T_{\rm c}$ with small oscillation and thus the dynamic regime prevails with small oscillations in $\dot{\varphi},i_{\rm wl},i_{\rm sh}$ as well, see Fig. \ref{fig5} for a $i_{\rm b}=1.68$. This regime exists until $i_{\rm b}=i_{\rm sr}$ ($i_{\rm sr}=2.449$) beyond which the WL goes to fully resistive state with temperature $T\geq T_{\rm c}$, \mbox{i.e.}, $p\geq 1$.

\begin{figure}[h!]
	\centering
	\includegraphics[width=\columnwidth]{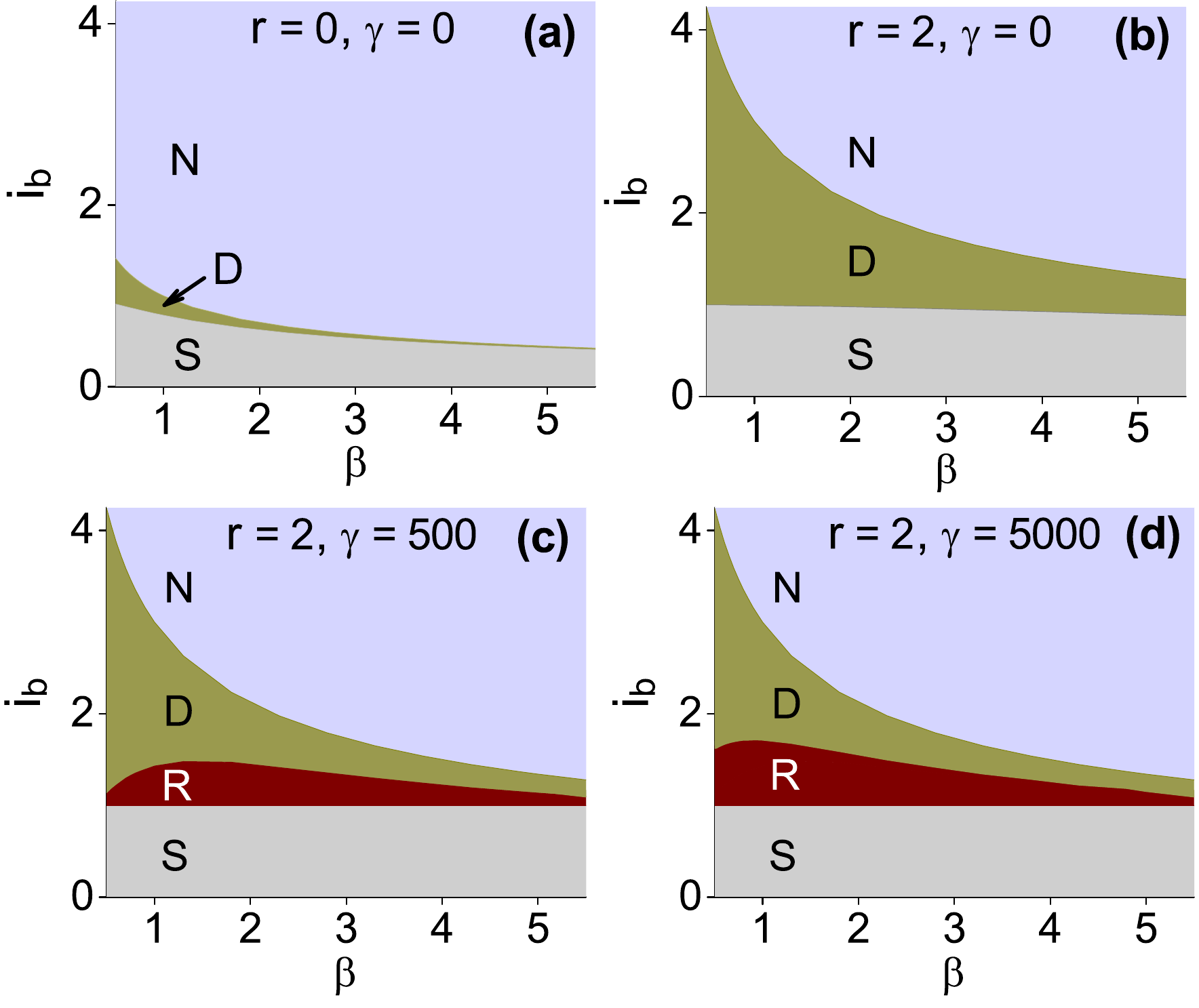}
	\caption{State diagrams describing distinct regimes for (a) $r=0$, $\gamma=0$, \mbox{i.e.}, no shunt, (b) $r=2$, $\gamma=0$, (c) $r=2$, $\gamma=500$ and (d) $r=2$, $\gamma=5000$. S, R, D, N represent the steady superconducting, relaxation, dynamic oscillatory and steady normal regime, respectively.}	
	\label{fig6}
\end{figure}	
\begin{figure}[h!]
	\centering
	\includegraphics[width=0.8\columnwidth]{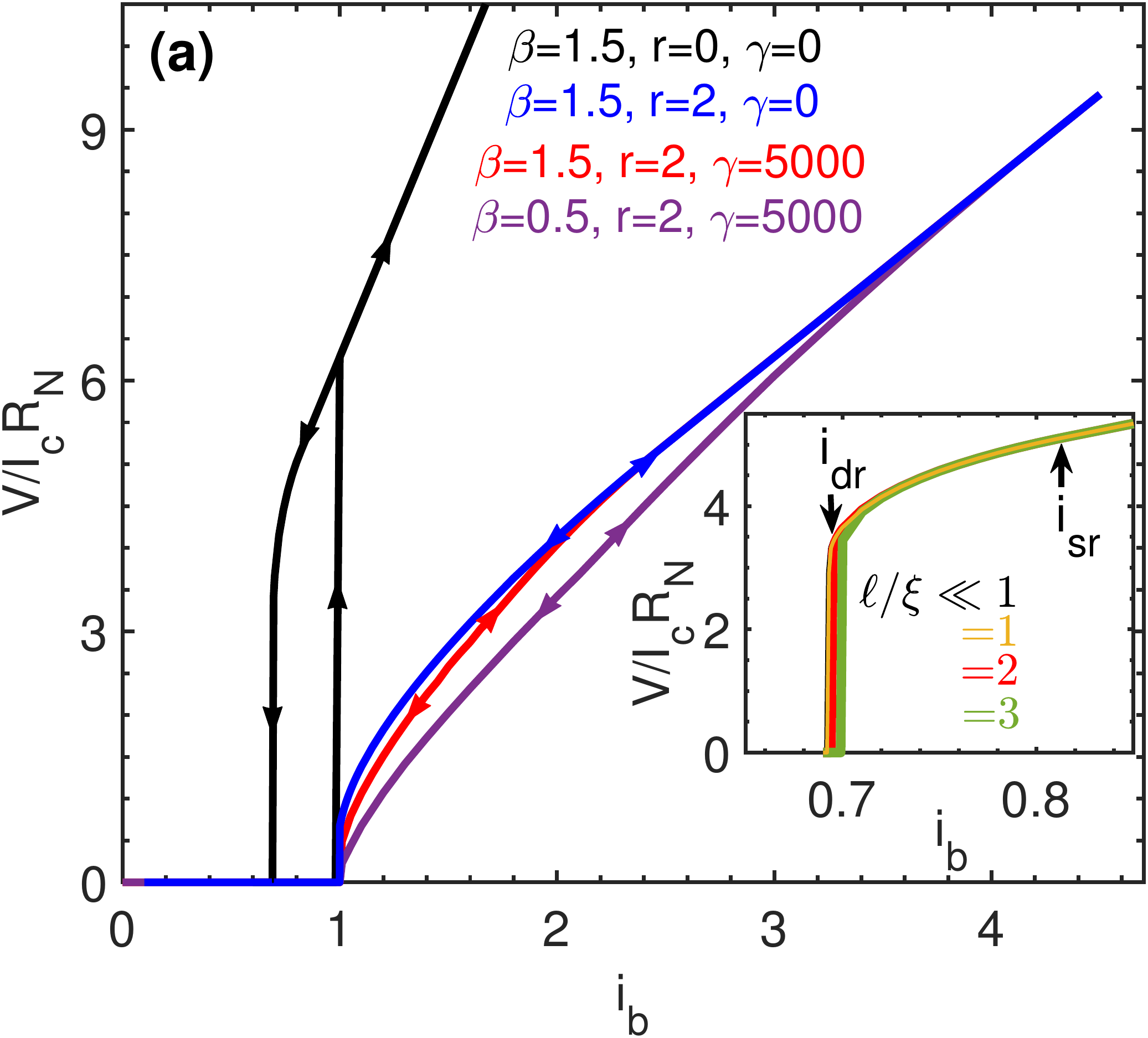}
	\includegraphics[width=0.8\columnwidth]{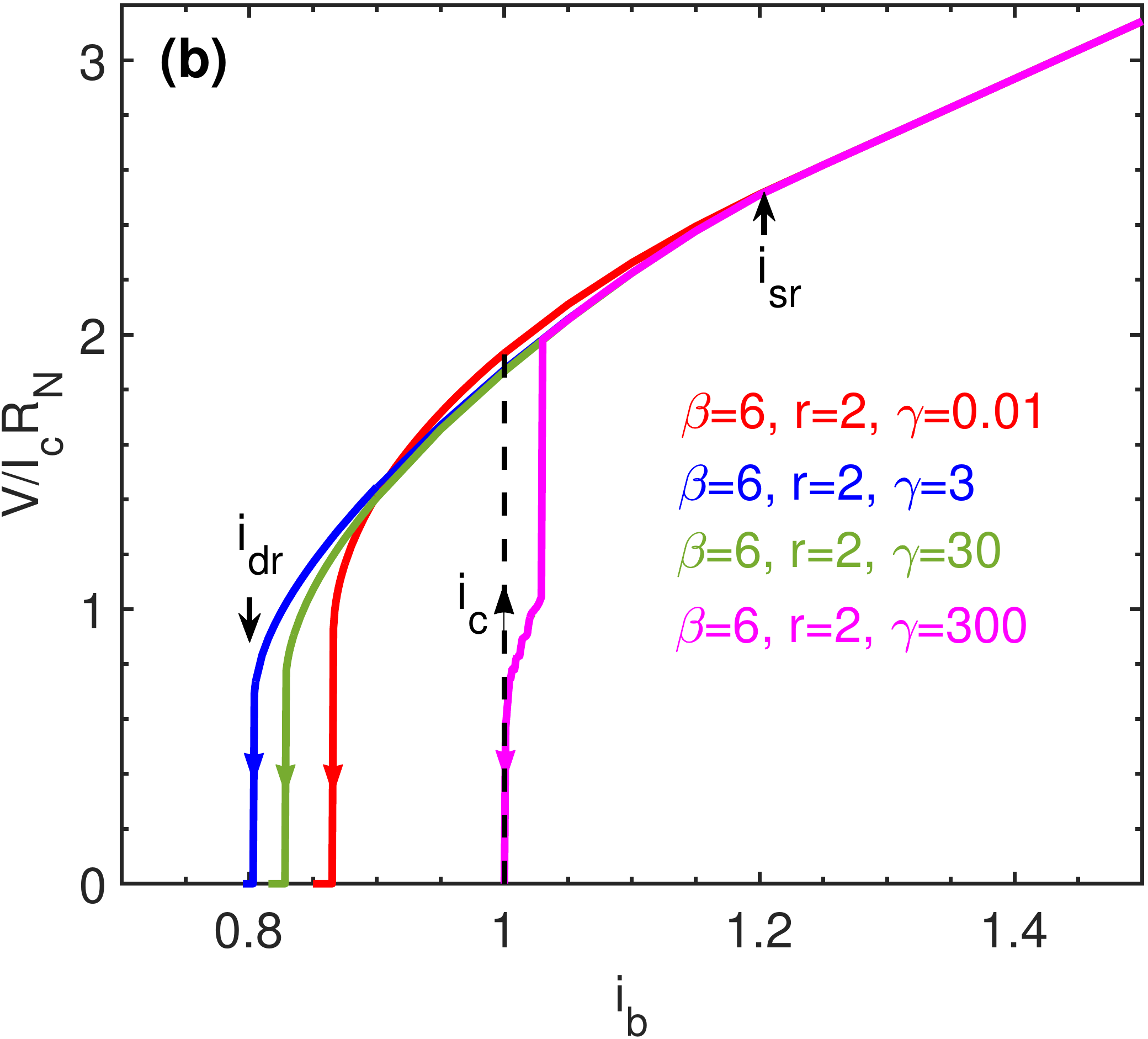}
	\caption{(a) DC current-voltage characteristics (IVCs) of the WL for different parameter sets covering all distinct regimes. Inset shows the branch of IVCs from the dynamic ($i_{\rm dr}$) to static ($i_{\rm sr}$) retrapping current for sinusoidal and non-sinusoidal supercurrent-phase relation with $r=0$ (no shunt case) at $\beta=1.5$. (b) IVCs for $\beta=6$, $r=2$ at four $\gamma$ showing the impact of $\gamma$ on the hysteresis. $\gamma$ values for blue, red, green and magenta curves correspond to the $i_{\rm dr}$-minimum, one from its left plateau, one from right plateau and one just below $i_{\rm dr}=1$ of the Fig. \ref{fig3} green curve, respectively.}
	\label{ivc}	
\end{figure}
We thus obtain various WL states, namely, steady superconducting (S), relaxation (R), dynamic oscillatory (D), and steady normal (N). Figure \ref{fig6} displays all these state diagrams for a few sets of parameters. Incorporation of the resistive shunt with no inductance ($L$) widens the dynamic oscillatory regime \big[Fig. \ref{fig6}(a) and \ref{fig6}(b)\big]. In presence of large $L$, a steady relaxation regime appears before the dynamic regime \big[Fig. \ref{fig6}(c)]. The higher the $\gamma$ is, the wider is the relaxation regime \big[Fig. \ref{fig6}(d)]. The latter shrinks with increased $\beta$, as a higher value of $\beta$ generates more heat than evacuation, which restricts the critical current $i_{\rm c}(p)$ to increase much and hence to cross the WLs' normal current $i_{\rm wl}$ before saturation. Note that a stable relaxation oscillation will arise only when the combination of $\beta$ and $r$ has to be such that the static retrapping current $i_{\rm sr}$ is greater than $1$.

Figure \ref{ivc}(a) shows the DC IVCs corresponding to all the WLs' states at $\beta=1.5$ and $0.5$. The DC voltage $V$ is calculated from the time-averaged value of $\dot{\varphi}$. Note that with shunt ($r=2$), the hysteresis almost disappears and there is a significant reduction in the voltage. As mentioned earlier, we also plot the IVCs for different $g(\varphi)$ for a $\beta=1.5$ and $r=0$ by numerically solving Eqs. (\ref{eq:eq2}) and (\ref{eq:eq4}), see the inset of Fig. \ref{ivc}. Here, the hysteretic IVCs almost merge except a negligible shift in the dynamic retrapping current $i_{\rm dr}$. We see a maximum of $0.8\%$ deviation in $i_{\rm dr}$ for $g(\varphi)$ corresponding to $\ell/\xi=3$ as compared to that for sinusoidal $g(\varphi)$.

To obtain IVCs over the intermediate $\gamma$ range, we take a large value of $\beta=6$ (for strongly hysteretic WL). Figure \ref{ivc}(b) presents the plotted IVCs for different $\gamma$ depicting how the hysteretic regime is affected by $\gamma$. The $\gamma$ values are chosen from the distinct regimes of Fig. \ref{fig3}, namely, one corresponding to the $i_{\rm dr}$-minimum, its left plateau, right plateau, and one corresponding to $i_{\rm dr}\approx1$.

\section{Discussion and Conclusion}	
Incorporating the relevant device- and thermal parameters, this model describes all dynamical perspectives of a Josephson WL. It is a more general thermal model which captures both shunted and unshunted WL characteristics via the state diagrams. For experiments with such Josephson devices, this model can provide controlled guidance.

A WLs' thermal state is described by the single parameter $\beta$ which is estimated from the retrapping current \cite{sourav}. At low temperatures, $\beta$ of a WL often is high making it hysteretic with a vanishing dynamic regime. Such a state of the WL is useful for studying switching statistics and in single-photon detection \cite{singlephoton}. A lesser $\beta$ is preferable for non-hysteretic WL and nano-SQUID application as voltage transducer. An optimally designed WL can have less $\beta$; still far from ideal for nano-SQUID. A parallel resistor and an inductor, chosen suitably, lead to a highly sensitive nano-SQUID \cite{souravRL,sagarnano}. Tuning the bias current and/or the shunt inductance, the WL can be probed in a distinct regime of relaxation oscillation.

In conclusion, we presented a dynamic thermal model by considering unavoidable Joule heating in the AC Josephson regime of a current biased resistively and inductively shunted superconducting WL. Our model captures various steady states of the WL, that are observed in the experiments. The state diagrams along with the IVCs are also discussed. This study will be important for a WL-based device to operate as a voltage read-out or a single-photon detector or a high-frequency relaxation oscillator. Also, our thermal model can be of interest to show a direction of new insights for graphene-, CNT-based JJ, and topological JJ.

\section*{Declaration of Competing Interest}
The authors declare that they have no known competing financial interests or personal relationships that could have appeared to influence the work reported in this paper.
\section*{Acknowledgements}
We acknowledge discussions with H. Courtois and financial support from the Council of Scientific and Industrial Research (CSIR) and the Department of Science and Technology (DST) of the Government of India.


\begin{thebibliography}{9}
	\bibitem{squidbook} J. Clarke and A. I. Braginski (Editors), \textit{The SQUID Handbook}, 2004.
	\bibitem{tinkham book} M. Tinkham, Introduction to Superconductivity 2nd ed. (Mc. Graw-Hill, New York, 1996).
	\bibitem{quantum} M. K. Devoret and R. J. Schoelkopf, Science {\bf339}, 1169 (2013).
	\bibitem{likharev} K. K. Likharev, Rev. Mod. Phys. {\bf51}, 101 (1979).
	\bibitem{granatasquid} C. Granata and A. Vettoliere, Phys. Rep. {\bf614}, 1-69 (2016).
	\bibitem{recentsquidreview} M. J. Mart\'inez-P\'erez and D. Koelle, \textit{NanoSQUIDs: Basics \& recent advances}, Phys. Sci. Rev. {\bf 2}, 20178001 (2017).
	\bibitem{tinkham}M. Tinkham, J. U. Free, C. N. Lau, and N. Markovic, Phys. Rev. B \textbf{68}, 134515 (2003).
	\bibitem{herve} H. Courtois, M. Meschke, J. T. Peltonen, and J. P. Pekola, Phys. Rev. Lett. {\bf 101}, 067002 (2008).
	\bibitem{anjanjap} A. K. Gupta, N. Kumar, S. Biswas, J. App. Phys. \textbf{116}, 173901 (2014).
	\bibitem{skocpol} W. J. Skocpol, M. R. Beasley, and M. Tinkham, J. App. Phys. \textbf{45}, 4054(1974).
	\bibitem{hazra} D. Hazra, L. M. A. Pascal, H. Courtois, and A. K. Gupta,  Phys. Rev. B {\bf82}, 184530 (2010).
	\bibitem{nikhil prl} N. Kumar, T. Fournier, H. Courtois, C. B. Winkelmann and A. K. Gupta, Phys. Rev. Lett. {\bf 114}, 157003 (2015).
	\bibitem{nayana} N. Shah, D. Pekker, and P. M. Goldbart, Phys. Rev. Lett {\bf 101}, 207001 (2008).
	\bibitem{singlephoton} A. J. Kerman, J. K. W. Yang, R. J. Molnar, E. A. Dauler, and K. K. Berggren, Phys. Rev. Lett. {\bf 79}, 100509(R) (2009).
	\bibitem{singlephoton-2} G. Oelsner et. al., Appl. Phys. Lett. {\bf 103}, 142605 (2013).
	\bibitem{singlephoton-new} L. S. Revin et. al., Beilstein J. Nanotechnol. {\bf 11}, 960 (2020).
	\bibitem{sourav} S. Biswas, C. B. Winkelmann, H. Courtois, and A. K. Gupta,  Phys. Rev. B {\bf 98}, 174514 (2018).
	\bibitem{nanomag1} W. Wernsdorfer, Adv. Chem. Phys. {\bf118}, 99 (2001).
	\bibitem{nanomag2} C. Veauvy, K. Hasselbach, and D. Mailly, Rev. Sci. Instrum. {\bf73}, 3825 (2002).
	\bibitem{nanomag3} D. Vasyukov et. al., Nature Nanotech. {\bf8}, 639 (2013).
	\bibitem{lam} S. K. H. Lam and D. L. Tilbrook, Appl. Phys. Lett. {\bf 82}, 1078 (2003).
	\bibitem{dibyendu} D. Hazra, J. R. Kirtley, and K. Hasselbach, Appl. Phys. Lett. {\bf 103}, 093109 (2013).
	\bibitem{nikhil sust} N. Kumar, C. B. Winkelmann, S. Biswas, H. Courtois and A. K. Gupta, Supercond. Sci. and Technol. {\bf 28}, 072003 (2015).
	\bibitem{souravRL} S. Biswas, C. B. Winkelmann, H. Courtois, T. Dauxois, H. Biswas, and A. K. Gupta, Phys. Rev. B {\bf 101}, 024501 (2020).
	\bibitem{sagarpra} S. Paul, G. Kotagiri, R. Ganguly, H. Courtois, C. B. Winkelmann and A. K. Gupta, Phys. Rev. Appl. {\bf 15}, 024009 (2021).
	\bibitem{muck} M. M\"uck, H. Rogalla, C. Heiden, Appl. Phys. A \textbf{46} 97-101(1988).
	\bibitem{zimmerman} J. E. Zimmermann and A. H. Silver, Phys. Rev. Lett. {\bf19}, 14 (1967).
	\bibitem{taur} Y. Taur and P. L. Richards, J. Appl. Phys. {\bf46}, 1793 (1975).
	\bibitem{vernon} F. L. Vernon, and R. J. Pedersen, J. Appl. Phys. {\bf39}, 2661 (1968).
	\bibitem{relaxapl} N. Calander, T. Claeson and S. Rudner, Appl. Phys. Lett. {\bf39}, 504 (1981).
	\bibitem{relaxiop} N. Calander, T. Claeson and S. Rudner, Physica Scripta {\bf25}, 837 (1982).
	\bibitem{sullivan} Sullivan, D. B., Peterson, R. L., Kose, V. E. and Zimmerman, J. E., J. Appl. Phys. {\bf41}, 4865 (1970).
	\bibitem{whan} C. B. Whan, C. J. Lobb and M. G. Forrester, J. Appl. Phys. {\bf77}, 382 (1995).
	\bibitem{calander} N. Calander, T. Claeson and S. Rudner, Appl. Phys. Lett. {\bf53}, 5093 (1982).
	\bibitem{kerl berggren mit} E. Toomey, Qing-Yuan Zhao, A. N. McCaughan and Karl K. Berggren, Phys. Rev. Appl. {\bf 9}, 064021 (2018).
	\bibitem{hadfield} R. H. Hadfield, A. J. Miller, S. W. Nam, R. L. Kautz and
	R. E. Schwall, Appl. Phys. Lett. {\bf 87}, 203505 (2005).
	\bibitem{relax-application1} B. van der, P. D. Sc, and J. van der Mark, LXXII. Lond. Edinb. Dublin Philos. Mag. J. Sci. {\bf6}, 763 (1928).
	\bibitem{relax-application2} R. Lang and K. Kobayashi, IEEE J. Quantum Electron. {\bf12}, 194 (1976).
	
	\bibitem{wolf} E. de Wolf and R. de Bruyn Ouboter, Physica B. {\bf176}, 133 (1992).
	\bibitem{golubov} A. Golubov et. al., Rev. Mod. Phys. {\bf76}, 411 (2004).
	\bibitem{whan_chaos} A. B. Cawthorne, C. B. Whan and C. J. Lobb, J. Appl. Phys. {\bf84}, 1126 (1998).
	\bibitem{sagarnano} S. Paul, G. Kotagiri, R. Ganguly, A. Subramanian, H. Courtois, C. B. Winkelmann, and A. K. Gupta, Phys. Rev. B {\bf 105}, L180410 (2022).
	
	
\end{thebibliography}
\end{document}